\documentclass{moriond}

\usepackage{comment}
\usepackage{wrapfig}
\usepackage{amsmath,amssymb}
\def\be{\begin{equation}}
\def\ee{\end{equation}}
\def\bea{\begin{eqnarray}}
\def\eea{\end{eqnarray}}

\begin{document}
\vspace*{4cm}
\title{Gravitational wave cosmology with extreme mass-ratio inspirals}

\author{\vspace*{-1mm}Danny Laghi}

\address{\vspace*{1mm}Dipartimento di Fisica ``Enrico Fermi'', Universit\`a di Pisa, and INFN Sezione di Pisa,\\ I-56127 Pisa, Italy}

\maketitle\abstracts{
We show that the loudest extreme mass-ratio inspirals (EMRIs) detected by the future space-based gravitational wave detector LISA can be used as dark standard sirens, statistically matching their sky localisation region with mock galaxy catalogs. 
In these Proceedings we focus on a realistic EMRI population scenario and report accuracy predictions for the measure of cosmological parameters, anticipating the potential of EMRIs to simultaneously constrain the Hubble constant, the dark matter, and the dark energy density parameters.}

\section{Introduction}

The Laser Interferometer Space Antenna (LISA) promises access to the mHz frequency window of the gravitational wave (GW) spectrum~\cite{LISA}.
Extreme mass-ratio inspirals (EMRIs) are inspiralling binary black hole systems formed by a massive black hole ($\sim$10$^6 M_{\odot}$) and a smaller compact object ($\sim$10$M_{\odot}$) that will be detected by LISA. 
In these Proceedings we assume a realistic EMRI population scenario and give: i) accuracy predictions for $h \equiv H_0/100$ km$^{-1}$ s Mpc and $w_0$, the dark energy (DE) equation-of-state parameter, for both 4 and 10 years of LISA mission~\cite{Laghi:2021pqk}; ii) \emph{preliminary} accuracy predictions for $h$ jointly inferred with the density parameters $\Omega_m$ and $\Omega_{\Lambda}$ (or equivalently $\Omega_{k}$) for 10 years of LISA mission.
Our method is based on the joint use of these \emph{dark sirens} measured distance along with their redshift, obtained by statistically matching~\cite{Schutz} their sky localisation region with mock galaxy catalogs. 
We show that high signal-to-noise-ratio (SNR) EMRIs are well-suited for doing cosmological Bayesian inference with dark sirens.

\section{EMRI \& Galaxy  Catalogs}

We use EMRI catalogs~\cite{Babak:2017tow} which assume 10 years of LISA mission lifetime with an arm-length baseline of $\sim$2.5$\times 10^6$ km, while we use galaxy catalogs~\cite{2012MNRAS.421.2904H} which are based on the Millennium Simulation~\cite{2005Natur.435..629S}. We assume as nominal \emph{true} cosmology the Millennium one, i.e.,~$h=0.73$, $\Omega_m=0.25$, $\Omega_{\Lambda}=0.75$. Here we focus on a \emph{fiducial} EMRI scenario (in a dedicated study~\cite{Laghi:2021pqk} we have also investigated more pessimistic and optimistic scenarios, the grade of confidence reflecting the number of available events).
To each EMRI we associate a true and potential hosts drawn from the galaxy catalog, that we use up to $z<1$.
Possible hosts are any galaxies with stellar mass $M_*>10^{10}{\rm M}_{\odot}$ that lie within the 2$\sigma$ credible region for the GW direction and distance, for at least one value of the cosmological parameters within the prior range. This selects a 3D co-moving volume in direction and redshift containing $N_{g,i}$ possible hosts for the $i$'th EMRI, each host having sky position ($\theta_j,\phi_j$) and redshift $z_j$. 
To account for statistical fluctuations in the galaxy distribution, that we verified to be complete out to $z\sim0.5$, we repeat the host-drawing procedure three times for each EMRI scenario. We select and analyse only the most informative events with SNR$>$100.
In Fig. 1 we show the relevant properties of our fiducial EMRI model. 

\begin{wrapfigure}{r}{0.5\textwidth}
\vspace{-0.55cm}
    \begin{center}
        \includegraphics[width=0.9\linewidth]{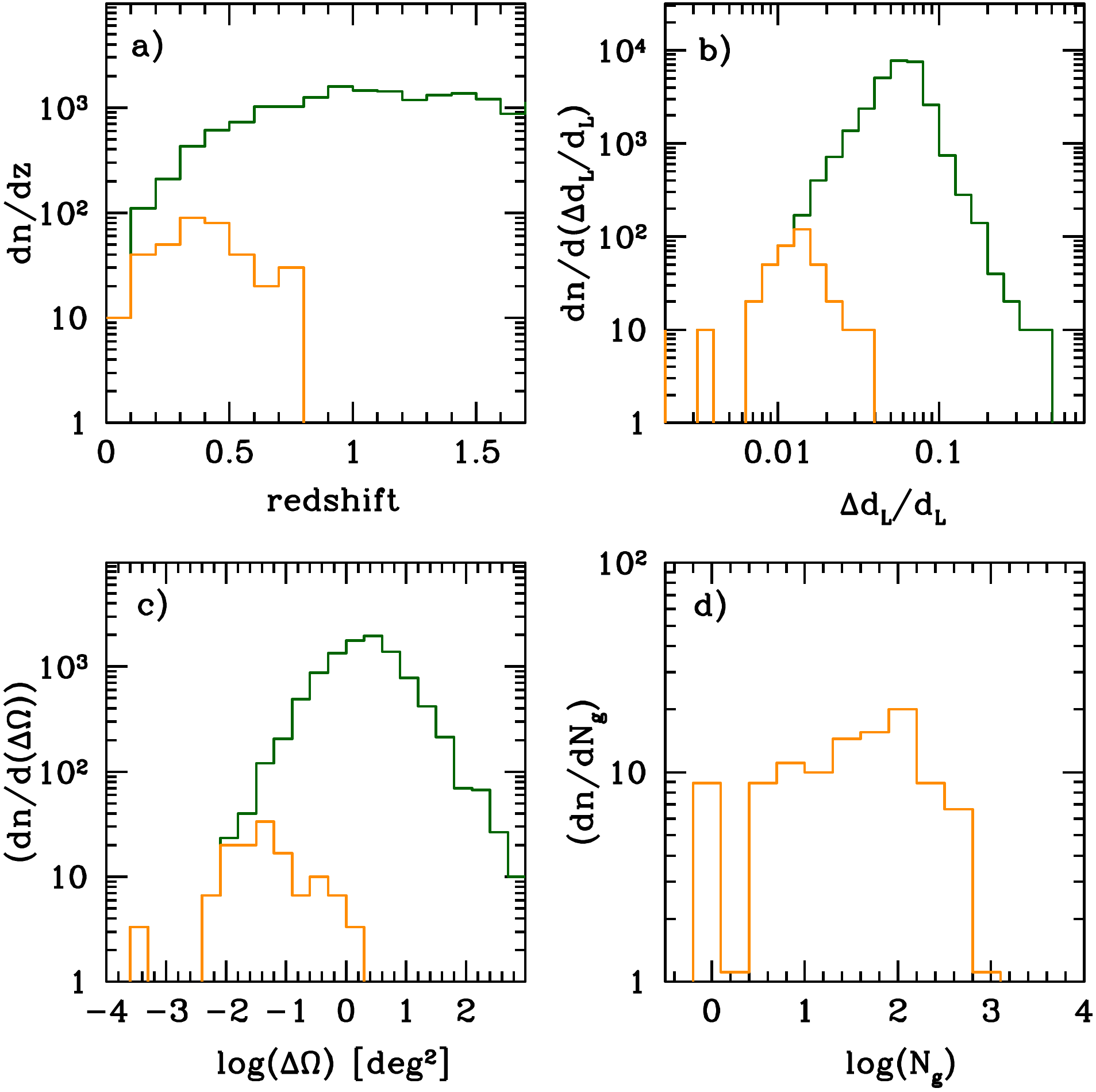}
    \end{center}
    \caption{Properties of our fiducial EMRI model. Green histograms in panels a), b), and c) show the redshift, luminosity distance error ($\Delta d_L/d_L$), and sky location error ($\Delta \Omega$) distributions, respectively, for the entire EMRI population detected by LISA in 10 years of operations ($\mathcal{O}(3000)$ with SNR$>$20). For the purpose of our analysis we select only systems with SNR$>$100, which are $N = \mathcal{O}(30)$. This selection criterion automatically selects the most informative events within $z\lesssim0.75$, i.e., well-localised events with relatively few hosts ($N_g\lesssim1000$). Orange histograms in panels a), b), and c) show these events. Panel d) shows the distribution of the number of candidate hosts within the 3D 2$\sigma$ error volume, averaged over the three realisations of the model.}
    \vspace{-0.3cm}
\end{wrapfigure}

\section{Bayesian Inference}

According to Bayes' theorem, the \emph{posterior} probability distribution for the cosmological parameters $\Omega$ is:
\begin{equation*}\label{eqn:bayes-cosmo}
    p(\Omega\,|\,D\,\mathcal{H}\,I) = p(\Omega\,|\,\mathcal{H}\,I)\,\frac{p(D\,|\,\Omega\,\mathcal{H}\,I)}{p(D\,|\,\mathcal{H}\,I)} \,,
\end{equation*}
where $\mathcal{H}$ is our cosmological model, which assumes a flat Friedmann-Lema\^itre-Robertson-Walker metric prescribing a luminosity distance-redshift functional relation, $d_L = d(\Omega, z)$, $I$ represents all the relevant background information, and $D\equiv \{ D_1,\ldots,D_N \}$ is the set of GW observations, with $D_i$ the data from the $i$'th EMRI event. 
The LISA 3D error volume is approximated by a multivariate Gaussian distribution:
\begin{equation*}
  p(D_i | {\bf \Theta})=\frac{1}{\sqrt{(2\pi)^3|{\bf \Sigma|}}} \exp\left\{\!-\dfrac{1}{2}{\bf \tilde{\Theta}}_i^{T}{\bf \Sigma}^{-1}{\bf \tilde{\Theta}}_i\right\} \text{,}
\label{eq:psky}
\end{equation*}
where ${\bf \tilde{\Theta}}_i \equiv {\bf \Theta}-\hat{\bf \Theta}_i(D_i)$, and 
the 3D parameter vector includes the source luminosity distance, the cosine of its declination, and its right ascension, i.e., ${\bf \Theta}=(d_L,\cos\theta_{gw}, \phi_{gw})$.
The vector ${\bf \hat{\Theta}}_i(D_i)=(\hat{d}_L,\cos\hat{\theta}_{gw}, \hat{\phi}_{gw})_i$ defines the best measured values of the parameters.
${\bf \Sigma}$ is the 3D correlation matrix, which we extract from the full 15D correlation matrix of the EMRI parameter estimation~\cite{Babak:2017tow}.
We expect the observed events to have almost zero correlation with each other; thus the likelihood function can be written as:
\begin{equation*}
    p(D\,|\,\Omega\,\mathcal{H}\,I) = 
\prod_{i=1}^N p(D_i\,|\,\Omega\,\mathcal{H}\,I)\text{.}
\end{equation*}
Assuming that correlations between the LISA measurements of the sky position angles and distance can be neglected, the single-event likelihood that we use is given by:
\begin{equation*}
\begin{split}
 p(D_i\,|\,\Omega\,\mathcal{H}\,I) =  \,
     &\frac{1}{2\pi} \int   \textup{d}z_{gw,i} 
         \sum_{j=1}^{N_{g,i}} 
    \frac{w_j}{\sigma_{z_j} \sqrt{\sigma_{\hat{d}_{L,i}}^2\!+\sigma_{W\!L,i}^2}} \times\\
    & \exp{\Biggl\{\! -\frac{1}{2} \Biggl[ \frac{\bigl( z_j-z_{gw,i} \bigr)^2}{\sigma_{z_j}^2} +\frac{\bigl(\hat{d}_{L,i} - d(\Omega, z_{gw,i})\bigr)^2}{\sigma_{\hat{d}_{L,i}}^2\!+\sigma_{W\!L,i}^2} \Biggr] \Biggr\}}
\,,
\end{split}
\end{equation*}
where $w_j$ are weights computed from $p(D_i | {\bf \Theta})$ marginalised over $d_L$ and evaluated at the sky position angles ($\theta_j,\phi_j$) of the galaxy $j$. We account for the galaxy peculiar velocity uncertainty with $\sigma_{z_j}$~\cite{DelPozzo:2017kme}, while including the weak lensing contribution $\sigma_{W\!L,i}$ in the distance measure uncertainty~\cite{Tamanini:2016zlh}. We verified that both selection and completeness effects do not significantly affect our conclusions.
We explore the posterior distribution $p(\Omega\,|\,D\,\mathcal{H}\,I)$ using \texttt{cpnest}~\cite{cpnest}, a parallel nested sampling algorithm implemented in~\texttt{Python}. The inference code utilised in this paper is publicly available in a dedicated \texttt{GitHub} repository~\cite{cosmolisa}.

\section{Fiducial EMRI Model}

In these Proceedings, we focus on our fiducial EMRI population model, assuming both 4 and 10 years of LISA observation; we consider two cosmological scenarios:

i) a $\Lambda$CDM scenario, characterized by a parameter space $\Omega \equiv \{h,\Omega_m \}$ to be explored, assuming a uniform prior range $h\in[0.6,0.86]$ and $\Omega_m\in[0.04,0.5]$, with fiducial values dictated by the Millennium run;

ii) a DE scenario, in which we assume $(h,\Omega_m,\Omega_{\Lambda})$ to be pre-determined by other probes at the values of the Millennium run and we search on the parameters $\Omega \equiv \{w_0,w_a\}$ defining the DE equation of state $w(z) = w_0 + z/(1+z) w_a$, drawing from the uniform prior range $w_0\in[-3,-0.3]$, $w_a\in[-1,1]$, with fiducial values $w_0 = -1$ and $w_a = 0$, corresponding to the cosmological constant $\Lambda$. 

For the 10-year analysis, posteriors are averaged over the analyses of the three different 10-year realisations, while for the 4-year case we draw three sets of events according to a Poisson distribution from each 10-year realisation, for a total of nine 4-year realisations, which then we separately analyse and average.
Assuming a ten (four) year LISA mission, $H_0$ and $w_0$ are constrained to $\sim$1.5\% ($\sim$2.5\%) and $\sim$6.2\% ($\sim$7.4\%), respectively, within 90\% credible regions (CRs).
Our $\Lambda$CDM results can be easily illustrated with a $d_L$-$z$ regression line, as in Fig.~2, which is obtained from one of the three 10-year EMRI realisations.
\begin{figure}[h]
\begin{center}
    \includegraphics[width=0.99\linewidth]{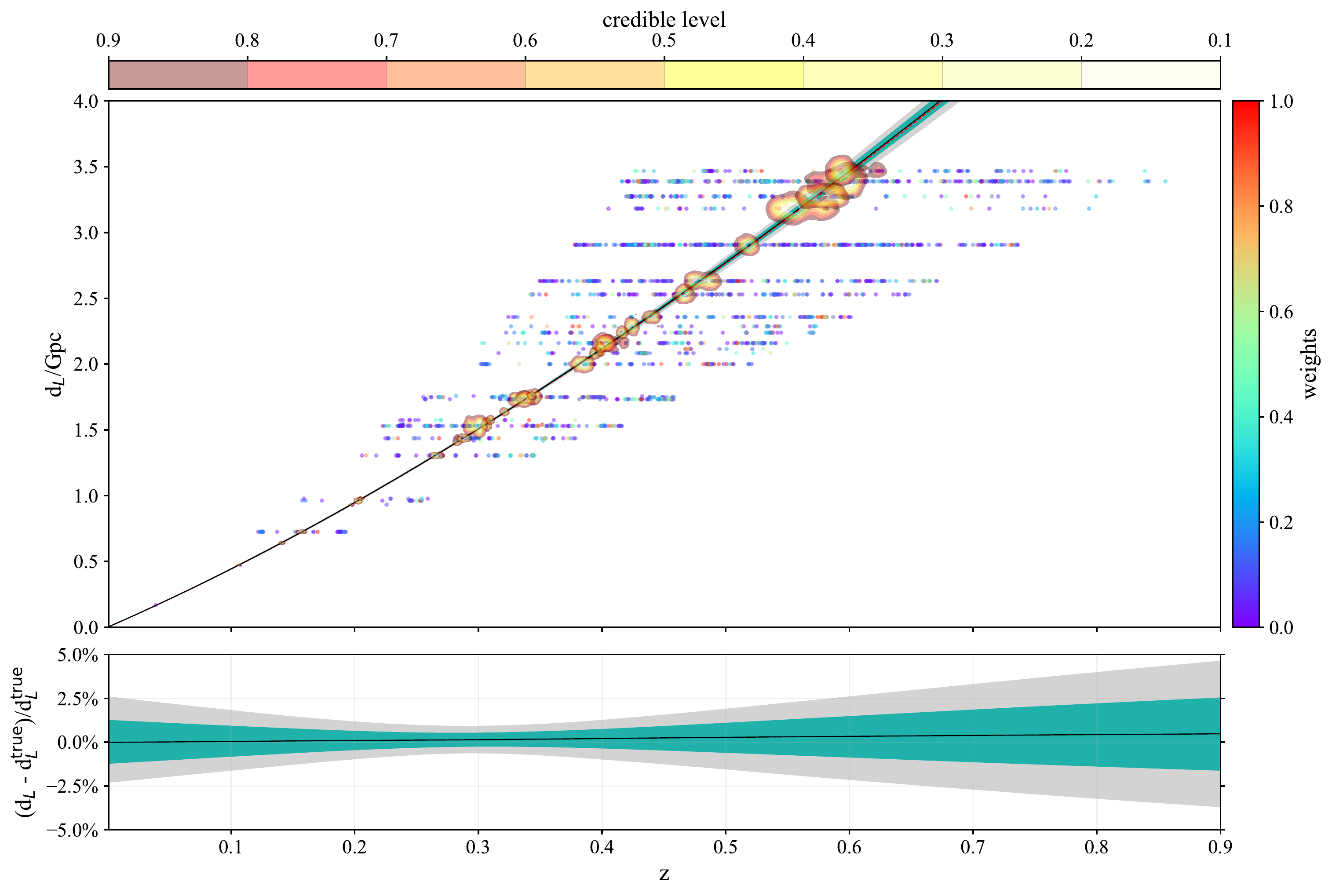}
    \caption{Regression line for one of the three 10-year realisations of our fiducial EMRI model. The median (solid black) and 68\% and 95\% CRs are in light seagreen and light gray, respectively. The red dashed line corresponds to the Millennium Simulation fiducial values of $h$, $\Omega_m$, and $\Omega_{\Lambda}$. The coloured blobs show the posterior distribution for $z_{gw}$ and $d_L$ for each EMRI event. The horizontal dots show the redshift of each candidate galaxy host for that particular EMRI. For illustrative purpose, we assigned to each galaxy a luminosity distance equal to $\hat{d}_L$. The dots are colour-coded from violet to red for increasing values of the weights $w_j$. The bottom panel shows the residuals of the inferred regression line CRs.}
\end{center}
\end{figure}

\section{Full $\Lambda$CDM: Future Prospects}

\begin{wrapfigure}{r}{0.5\textwidth}
\vspace{-0.7cm}
    \begin{center}
        \includegraphics[width=0.99\linewidth]{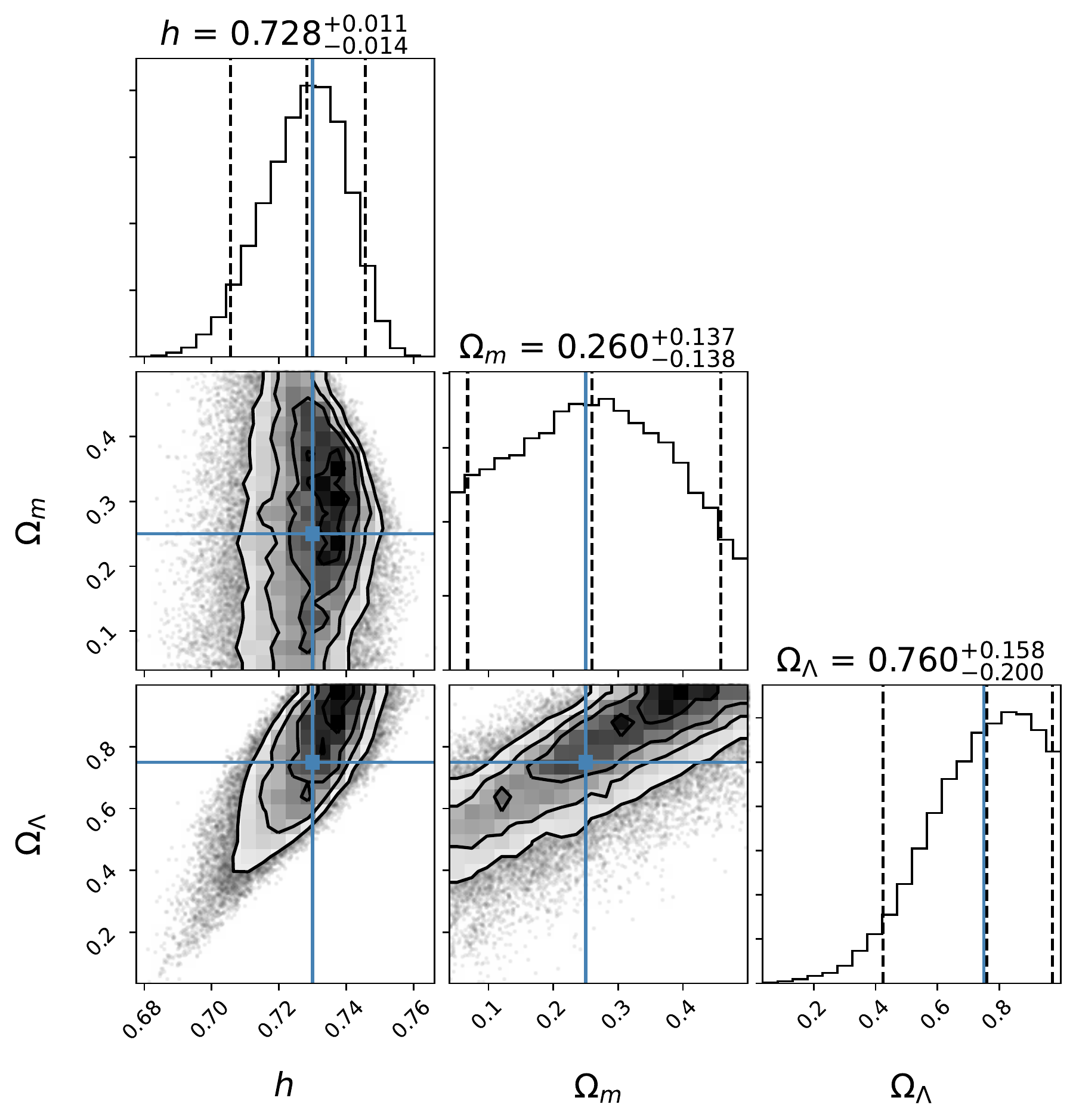}
    \end{center}
    \caption{Posterior distributions obtained by sampling the full $\Lambda$CDM cosmological parameter space, $\Omega \equiv \{h,\Omega_m,\Omega_{\Lambda} \}$, assuming our fiducial EMRI scenario. In each panel, the cyan lines mark the fiducial values and the black dashed lines indicate the median and 90\% CRs extracted from the marginalized posterior.}
    \vspace{-0.85cm}
\end{wrapfigure}
Preliminary investigations of the full $\Lambda$CDM scenario, where in addition to the uniform $h$-$\Omega_m$ prior defined above we assume a uniform prior range $\Omega_{\Lambda} \in [0,1]$, indicate that, even for moderate-redshift sources such as our loudest EMRIs at $z\lesssim 0.7$, LISA will provide simultaneous constraints on all cosmological parameters. This is shown for our fiducial EMRI scenario in Fig.~3.
Preliminary results (90\% CR) indicate that, in a 10-year LISA mission, $\Omega_{\Lambda}$ (or the curvature density $\Omega_k$) could be constrained with an accuracy of $\sim$27\% and $\Omega_m$ with an accuracy of $\sim$55\%, while retaining an accuracy on $H_0$ of $\sim$2\%.
This suggests that the joint inference of all the $\Lambda$CDM parameters with LISA standard sirens will be possible.

\section{Summary}

By using the loudest EMRIs detected by LISA (SNR$>$100) as dark standard sirens, we find that constraints on $H_0$ can reach $\sim$1.1\% ($\sim$3.6\%) accuracy (90\% CR), in our best (worst) case scenario~\cite{Laghi:2021pqk}.
By considering a dynamical DE equation of state, with $\Lambda$CDM parameters fixed by other observations, we further show that in our best (worst) case scenario $\sim$5.9\% ($\sim$12.3\%) relative uncertainties (90\% CR) can be obtained on $w_0$~\cite{Laghi:2021pqk}. Here we also suggest that EMRIs can be potentially used to constrain the $\Lambda$CDM model \emph{beyond} $\Omega_m$.
These results update the only EMRI cosmological analysis available in the literature to date~\cite{MacLeod:2007jd}.
EMRI measurements will be affected by different systematics compared to both electromagnetic and ground-based GW observations, hence they will provide additional independent validation of our present understanding of the Universe.

\section*{References}
\begin{small}

\end{small}


\end{document}